\documentclass[aps,pra,reprint,twocolumn,showpacs,superscriptaddress]{revtex4-1}

\usepackage{amsmath, amssymb, braket, dsfont, units}

\usepackage[mathscr]{euscript}
\usepackage{graphicx}
\usepackage{subfigure}		
\usepackage{rotating, multirow}
\usepackage{epstopdf, times}

\usepackage[colorlinks=true,linkcolor=blue,citecolor=blue,breaklinks]{hyperref}
\usepackage{breakurl}
\usepackage{color}

\newcommand{\corrR}{{g}}
\newcommand{\corrM}{{\tilde g}}
\newcommand{\up}{{\uparrow}}
\newcommand{\dn}{{\downarrow}}
\renewcommand{\r}{{\mathbf{r}}}
\newcommand{\shift}{\mathscr{S}}
\newcommand{\D}{\mathrm{d}}
\newcommand{\DSAS}{\Delta_\textrm{SAS}}
\newcommand{\Dnorm}{\DSAS/\frac{e^2}{\epsilon\ell_B}}		
\newcommand{\barthird}{\ensuremath{\overline{1/3}}}
\newcommand{\Uone}{\ensuremath{\mathrm{U(1)}} }
\newcommand{\SU}[1]{\ensuremath{\mathrm{SU(#1)}}}
\newcommand{\spinon}{{\phi_s}}

\definecolor{Zcolour}{rgb}{0.992, 0.588, 0.22}
\definecolor{purple}{rgb}{0.5,0,0.5}

\begin{document}

\title{Competing  Abelian and non-Abelian topological orders in \texorpdfstring{$\nu = 1/3+1/3$}{filling 1/3+1/3} quantum Hall bilayers}
\author{Scott Geraedts}
\affiliation{Department of Physics and Institute for Quantum Information and Matter, California Institute of Technology, Pasadena, California 91125, USA}
\author{Michael P. Zaletel}
\affiliation{Department of Physics, Stanford University, Stanford, California 94305, USA}
\affiliation{Station Q, Microsoft Research, Santa Barbara, California 93106, USA}
\author{Zlatko Papi\'c}
\affiliation{Perimeter Institute for Theoretical Physics, Waterloo, Ontario N2L 2Y5, Canada}
\affiliation{Institute for Quantum Computing, Waterloo, Ontario N2L 3G1, Canada}
\author{Roger S. K. Mong}
\affiliation{Department of Physics and Institute for Quantum Information and Matter, California Institute of Technology, Pasadena, California 91125, USA}
\affiliation{Walter Burke Institute for Theoretical Physics, Pasadena, California 91125 USA}
\affiliation{Department of Physics and Astronomy, University of Pittsburgh, Pittsburgh, Pennsylvania 15260, USA}

\begin{abstract}
Bilayer quantum Hall systems, realized either in two separated wells or in the lowest two sub-bands of a wide quantum well, provide an experimentally realizable way to tune between competing quantum orders at the same filling fraction.
Using newly developed density matrix renormalization group techniques combined with exact diagonalization, we return to the problem of quantum Hall bilayers at filling $\nu = 1/3 + 1/3$.
We first consider the Coulomb interaction at bilayer separation $d$,  bilayer tunneling energy $\Delta_\textrm{SAS}$, and individual layer width $w$,  where we find a phase diagram which includes three competing Abelian phases:  a bilayer-Laughlin phase (two nearly decoupled $\nu = 1/3$ layers);  a bilayer-spin singlet phase;  and a bilayer-symmetric phase. 
We also study the order of the transitions between these phases.
A variety of non-Abelian phases have also been proposed for these systems.
While absent in the simplest phase diagram, by slightly modifying the interlayer repulsion we find a robust non-Abelian phase which we identify as the ``interlayer-Pfaffian" phase.
In addition to non-Abelian statistics similar to the Moore-Read state, it exhibits a novel form of bilayer-spin charge separation.
Our results suggest that $\nu = 1/3 + 1/3$ systems merit further experimental study.
\end{abstract}

\maketitle
\tableofcontents

\section{Introduction}

The remarkable experimental discovery of quantized resistance of a two-dimensional electron gas (2DEG) in strong perpendicular magnetic fields~\cite{Tsui-PhysRevLett.48.1559} has revealed many topologically ordered phases that form due to strong Coulomb interactions in a partially filled Landau level~\cite{prangegirvin}.
 Some examples include the 
 ``odd-denominator" fractional quantum Hall (FQH) states that belong to the sequence of Laughlin~\cite{Laughlin-PhysRevLett.50.1395}, hierarchy~\cite{Haldane1983, Halperin84} and ``composite fermion"~\cite{Jain:1989p294} states. 
 One of their prominent features is the presence of quasiparticles (``anyons") that carry fractional charges~\cite{Laughlin-PhysRevLett.50.1395} and obey fractional statistics~\cite{LeinaasMyrheim, Arovas-Schrieffer-Wilczek}. 
 More intriguing, ``non-Abelian" quasiparticles have been proposed to occur in several experimentally observed FQH states in the first excited Landau level. 
 Most notably, this is the case with an even-denominator filling factor $\nu=5/2$ state~\cite{Willett87}, believed to be described by the Moore-Read Pfaffian state~\cite{MooreRead, Greiter91, Greiter92} that contains non-Abelian anyons of the Majorana type~\cite{Nayak96, ReadGreen, BondersonGurarieNayak11}. 

The aforementioned hierarchies of Abelian and non-Abelian states are \emph{a priori} relevant when the FQH system can be described as a single partially occupied Landau level, that is, the electrons carry no internal degree of freedom.
However, ``multicomponent" FQH states are ubiquitous; most obviously electrons carry spin.
While the Coulomb energy scales as $e^2/\epsilon\ell_B [\unit{K}] \approx 50 \sqrt{B[\unit{T}]}$, assuming free electron values for the mass and $g$ factor in $\mathrm{GaAs}$, the Zeeman splitting is only $E_Z [\unit{K}] \approx 0.3 B[\unit{T}]$, suggesting that in many circumstances the ground state of the system may not be fully spin-polarized. 
Several classes of unpolarized FQH states have been formulated, including the so-called Halperin $(mmn)$ states~\cite{Halperin83} and spin unpolarized composite fermion states~\cite{Wu93, Wu94, Davenport12, Balram14}.
In materials such as $\mathrm{AlAs}$ or graphene, ordinary electron spin may furthermore combine with valley degrees of freedom, which can change the sequence of the observed integer and FQH states~\cite{Bishop07, Padmanabhan09, Gokmen10, Novoselov05, Zhang05, Du09, Bolotin09, Ghahari11, Dean11, Feldman12}. 

Here we study an important class of multicomponent FQH systems where the internal degrees of freedom correspond to a subband or layer index, generally referred to as \emph{pseudo-spin}. 
For example, if a 2DEG is confined by an infinite square well in the perpendicular $z$-direction, the effective Hilbert space may be restricted to several low-lying subbands of the quantum well (QW). 
In the most common case, the relevant subbands are the lowest symmetric and antisymmetric subbands of the infinite square well that play the role of an effective \SU2 degree of freedom. 
Furthermore, it is possible to fabricate samples that consist of two quantum wells separated by a thin insulating barrier. We refer to the latter type of device as the quantum Hall bilayer (QHB). 
The interest in bilayers and quantum wells comes from their experimental flexibility that allows one to tune the parameters in the Hamiltonian to a larger degree than it is possible with ordinary spin. For example, in a QHB with finite interlayer distance, the effective Coulomb interaction is not \SU2 symmetric. Therefore, the ``intralayer" Coulomb interaction (the potential between electrons in the same layer) is somewhat stronger than the ``interlayer" Coulomb (i.e., the potential between electrons in opposite layers).
The ratio between the two interaction strengths is given by the parameter $d/\ell_B$, the physical distance between layers in units of magnetic length, which in experiment can be continuously tuned.
The tunneling energy between the two layers (in units of the Coulomb interaction energy), $\Dnorm$, can also be tuned.
The tunability of interactions in quantum Hall bilayers and quantum wells can give rise to a richer set of FQH phases that extend beyond those realized in single-layer systems.
Examples of such phases occur at $\nu=1$ and $\nu=1/2$.
They have a rich experimental history that we briefly review in Sec.~\ref{sec:ex_back}. 

In this work we focus on the QHB at total filling factor $\nu=1/3+1/3$.
The early experiment by Suen \emph{et al.}~\cite{Suen94} measured the quasiparticle excitation gap in a wide QW as a function of $\DSAS$.
The gap was found to close around $\Dnorm \lesssim 0.1$, with an incompressible phase on either side of the transition.
A realistic model of this system~\cite{Lay97}, that included LDA calculation of the band structure, reproduced the observed behavior of the gap. 
A more complete phase diagram as a function of both $d/\ell_B$ and $\Dnorm$ was obtained in Ref.~\onlinecite{HaldaneDiagram}.
This study, however, assumed zero width for each layer and was restricted to small systems.
The phase diagram was argued to consist of three phases.
For small $d/\ell_B$ and small $\Dnorm$, the system maintains \SU2 symmetry and resembles the usual $\nu=2/3$ state with spin.
It has been known that the ground state in this case is a spin-singlet $(112)$ state~\cite{Chakraborty84, Rezayi87, Maksym89, Wu93} (for an explicit wavefunction see Refs.~\onlinecite{HaldaneDiagram,Moore97}). 
If $d/\ell_B$ is large, the layers are decoupled and the system is described by the Halperin $(330)$ state, which is the simple bilayer Laughlin state.
On the other hand, large $\DSAS$ effectively wipes out the layer degree of freedom, and the system becomes single component.
This bilayer symmetric state is described by the particle-hole conjugate of Laughlin's 1/3 wavefunction (hereafter called the $\barthird$ state).         

\begin{figure}[ttt]
	\includegraphics[width=\linewidth]{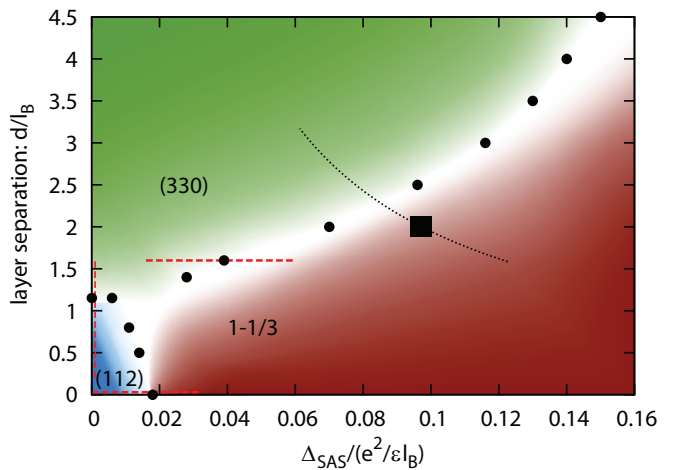}
	\caption{\label{phase} (Color online)
		Phase diagram of $1/3+1/3$ QHB in terms of dimensionless layer separation $d$ and tunnelling energy $\DSAS$. 
		Data was taken with cylinder circumference $L=14\ell_B$ and layer width $w=0$.
		The dashed lines indicate sweeps performed to determine the nature of the phase transitions (see Sec.~\ref{sec:phase} for details).
		Later in this work, additional axes will be added to this plot, driving the system into a non-Abelian phase (see Fig.~\ref{V0phase}).
		The black dashed line and square mark the region studied experimentally in Ref.~\onlinecite{Hirayama2002b}, and their observed phase transition.
	}
\end{figure}
Our motivation for revisiting the problem of $\nu=1/3+1/3$ QHB is twofold.
First, previous theoretical studies of this system have been limited to very small systems due to the exponential cost of exact diagonalization (ED). 
This limitation is particularly severe in the present case because of the pseudo-spin degree of freedom.
Recent work has demonstrated that to some degree this cost can be overcome by using variational methods such as the ``infinite density-matrix renormalization group" (iDMRG)~\cite{ZaletelQHdmrg13, ZaletelMixing}. 
By combining insights from ED and iDMRG, we are able to obtain a more accurate phase diagram of the $\nu=1/3+1/3$ QHB system as a function of $d$ and $\DSAS$, as shown in Fig.~\ref{phase}. 
Although our results are qualitatively consistent with Ref.~\onlinecite{HaldaneDiagram}, the access to significantly larger system sizes enables us to study the order of the associated phase transitions, which we find to be first order. 

Given that $1/3+1/3$ bilayer systems are experimentally available and allow a great deal of tunability (changing the layer width $w$, $d$ or $\DSAS$), our second goal is to explore the possibility of realizing more exotic (non-Abelian) phases in these systems by tweaking the interaction parameters. 
Indeed, recently a number of trial non-Abelian states have been proposed for these systems~\cite{Ardonne02, Barkeshli10, RezayiZ4, BondersonSlingerland, Barkeshli11, Barkeshli14, Vaezi14}.
At filling $\nu=1/3+1/3$, the relevant candidates are the $\mathbb{Z}_4$ Read-Rezayi state~\cite{Read-PhysRevB.59.8084}, the bilayer Fibonacci state~\cite{Vaezi14}, the ``intralayer-Pfaffian" and ``interlayer-Pfaffian" states~\cite{Barkeshli10}. 
The latter is an example of a spin-charge separated state and was first introduced in Ref.~\onlinecite{Ardonne02}.
We develop a diagnostic that detects spin-charge separation in the ground-state wavefunction using the entanglement spectrum. 
By varying the short-range Haldane pseudopotentials in the bilayer system at finite interlayer distance and tunneling, we find evidence for a non-Abelian phase that exhibits spin-charge separation and has non-trivial ground-degeneracy, consistent with the interlayer Pfaffian state. 
The phase is realized by either reducing the $V_0$ or increasing the $V_1$ pseudopotential component of the interaction, which may naturally occur as a consequence of strong Landau level mixing.    

The remainder of this paper is organized as follows.
In Sec.~\ref{sec:ex_back} we review some of the previous experimental work in QHB and QW systems.
In Sec.~\ref{sec:method} we introduce the model of the QHB and discuss the numerical methods and diagnostics for identifying the FQH phases and transitions between them. 
Sec.~\ref{sec:phase} contains our main results for the phase diagram of $1/3+1/3$ QHB as a function of parameters $w$, $d$ and $\DSAS$. 
We discuss in detail the three Abelian phases that occur in this system, and identify the nature of the transitions between them. 
In Sec.~\ref{sec:nonabelian} we explore the possible new phases when the interaction is varied away from the bare Coulomb point. 
We establish that the modification of short-range ($V_0$ or $V_1$) pseudopotentials leads to a robust non-Abelian phase that exhibits spin-charge separation and can be identified with the interlayer Pfaffian state.
Our conclusions are presented in Sec.~\ref{sec:conclusions}.

\section{Experimental Background}
\label{sec:ex_back}

In this Section we briefly review some of the important experiments on quantum Hall bilayers and wide quantum wells. 
As mentioned in the Introduction, one of the great advantages of studying these systems is the ability to experimentally tune parameters in the Hamiltonian, e.g., the interlayer separation and interlayer tunneling in a QHB.
Different samples can be constructed with different values for these quantities.
Tunneling energy is independent of layer separation since it can be varied by changing the height of the potential barrier between the layers without changing its width.
Another convenient way to tune these parameters is by applying voltage bias to separate contacts made to each layer~\cite{EisensteinSeparateContacts}; the variation of electron density $\rho$ thus changes the effective $\ell_B$ at fixing filling $\nu$ via the relation $\rho=\nu/2\pi\ell_B^2$.
This allows $d/\ell_B$ and $\Dnorm$ to be tuned continuously in a single sample. 

To illustrate the typical parameter range that can be accessed, we note that at $\nu=1/2+1/2$ it has been possible to vary $d/\ell_B$ in range 1.2\mbox{--}4, while the interlayer tunneling $\DSAS$ can be either completely suppressed or as large as $0.1e^2/\epsilon\ell_B$~\cite{JimReview}.
The width of individual layers in this case is less than $d$. 
On the other hand, in wide QWs one controls independently the width of the entire well and the tunneling amplitude $\DSAS$. 
The latter is defined as the energy splitting between the lowest symmetric and antisymmetric subbands, and typically varies between zero and $0.2e^2/\epsilon\ell_B$.
For systems where FQH can be observed, the physical width of the well is typically 30--\unit[65]{nm}~\cite{Shabani13}. 
Self-consistent numerical calculations estimate that this corresponds to an effective bilayer distance $d/\ell_B=3\mbox{--}7$, with individual layer widths $1.5\mbox{--}3\ell_B$~\cite{Shabani13}.    
The tunability via $d/\ell_B$ or $\Dnorm$ can engender new physics that does not arise in a single layer quantum Hall system.
Two important examples of such phenomena have been observed to occur at total filling factors $\nu=1/2$ and $\nu=1$.  
   
At total filling $\nu=1/2$, the QHB ground state is compressible in the limit of both very large and very small $d/\ell_B$.
At large $d/\ell_B$, it is described by two decoupled $1/4+1/4$ ``composite Fermi liquids"~\cite{HalperinLeeRead} (CFL), while around $d/\ell_B=0$ it is the spin unpolarized $1/2$ CFL. 
At intermediate $d/\ell_B$, an incompressible state forms when $d/\ell_B \lesssim 3$~\cite{SuenOneHalf, EisensteinOneHalf}. 
Numerical calculations performed over the years, primarily utilizing exact diagonalization~\cite{Chakraborty87, Yoshioka89, He93, Scarola01, Peterson10_2}, have confirmed that the incompressible state at vanishing interlayer tunneling is the Halperin 331 state~\cite{Halperin83}. 
More recently, there has been some renewed interest in the $\nu=1/2$ two component systems~\cite{Shabani09, Shabani13} due to the possible transition into the Moore-Read Pfaffian state as tunneling is increased~\cite{Nomura04, Papic10, Peterson10}. 
Analogous scenario may hold for QWs at filling $\nu=1/4$, where the competing phases are the Halperin (553) state and the 1/4 Pfaffian state~\cite{Papic09}.
Very recently, $\mathrm{GaAs}$ hole systems have been shown to realize an incompressible state at $\nu=1/2$ near the vicinity of Landau level crossing~\cite{Liu14}.      

As a second example of novel phases in QHB systems, we briefly mention the celebrated $\nu=1$ state (for recent reviews, see Refs.~\onlinecite{EisensteinMacDonald, JimReview}). 
At large $d/\ell_B$ the system is compressible (two decoupled CFLs), but undergoes a transition to an incompressible state for $d/\ell_B<2$, even at negligible interlayer tunneling.
The incompressible state is represented by the Halperin (111) state, which can also be viewed as a pseudo-spin ferromagnet~\cite{GirvinMacDonald}. 
This wavefunction encodes the physics of exciton superfluidity, with an associated Goldstone mode~\cite{Spielman00} and vanishing of Hall resistivity in the ``counterflow" measurement setup~\cite{Kellogg04, Tutuc04}. 
The existence of an incompressible state (consistent with an exciton superfluid) has been established in numerics~\cite{Schliemann01, Simon03, Shibata06, Moller08, Papic11}, though the questions about the details and nature of the transition, as well as the possibility of intermediate phases, remain open.

The case of total filling $\nu=2/3$, which is the subject of this paper, has been less studied compared to previous examples.
In the mentioned Ref.~\cite{Suen94} the transition between a one-component and two-component phase was detected as a function of $\DSAS$, while in Ref.~\cite{Lay97} similar data was obtained as a function of the tilt angle of the magnetic field.
These experiments have been performed on a single wide QW.
More recently, Refs.~\cite{Hirayama2002b} and \cite{Hirayama2004} have studied $\nu=1/3+1/3$ in a QHB sample which directly corresponds to the model we study. (see Sec.~\ref{sec:method})
By applying a voltage bias as described above, they perform four sweeps in the $d$, $\DSAS$ plane. 
In one sweep \cite{Hirayama2002b} they find a seemingly first-order transition at $d/\ell_B\approx2$, $\Dnorm\approx0.1$.
This sweep, and the location of the observed transition, are shown in Fig.~\ref{phase}.
Another sweep entirely in the large $\DSAS$ regime sees no phase transition, while two other sweeps are performed at small $\DSAS$.
These sweeps see a $\nu=2/3$ state at large $d/\ell_B$ which vanishes as the interlayer separation is decreased.
The rest of the phase diagram remains to be fully mapped out.
In our work we determine this phase diagram numerically, which can guide experiments towards realizing all the possible phases in this bilayer system.
Finally, we mention that very recently~\cite{Liu15} the stability of fractional quantum Hall states was investigated in a wide quantum well system with competing Zeeman and tunneling terms.
The Zeeman splitting was controlled by an in-plane magnetic field.
This system may not be fully captured by our model in Sec.~\ref{sec:method} because of the potentially strong orbital effect of an in-plane field in a wide QW.
It is possible, however, that the transition observed at $\nu=5/3$ in Ref.~\onlinecite{Liu15} is indeed in the universality class of $\barthird \to (112)$ transition that we identify in Sec.~\ref{sec:phase} below.

\section{Model and Method}\label{sec:method}

\subsection{The bilayer model}
We label the two layers of the bilayer with the index $\mu \in \{\up,\dn\}$, and consider  Hamiltonians of the general form
\begin{multline}
	H=\frac{1}{2} \int\! \D^2\r\, \D^2\r^\prime\, V_C^{\mu\nu}(\r-\r^\prime) n^{\mu}(\r)n^{\nu}(\r^\prime) \\
	- \frac{\DSAS}{2} \int\! \D^2\r\, c^{\mu\dagger}(\r) \sigma^x_{\mu \nu} c^{\nu}(\r), 
	\label{ham}
\end{multline}
where ${c^{\mu}}^\dagger (\r)$ creates an electron in layer $\mu$ at the position $\r\equiv(x,y)$. The first term is the Coulomb interaction, expressed in terms of the density operator 
\begin{eqnarray}\label{density}
	n^{\mu}(\r)=c^{\mu\dagger} (\r) c^{\mu}(\r).
\end{eqnarray}
for an electron in layer $\mu$. The precise form of the interaction term depends on the details of the bilayer. The second term encodes tunneling between the two layers.
When $V_C^{\mu\nu}$ is \SU2 symmetric this Hamiltonian is equivalent to a $\nu=2/3$ system with spin, and in this case $\DSAS$ can be thought of as the Zeeman splitting.

In Eq.~(\ref{ham}) we assumed that the perpendicular $z$ coordinate has been integrated out, leading to an effective two-dimensional Hamiltonian.
This is possible because the magnetic field is perpendicular to the 2DEG plane, and the transverse component of the single body wavefunctions $\psi$ factorizes,
\begin{equation}
\psi^\mu(x,y,z)=\phi_z(z \pm d / 2)  \phi(\r).
\end{equation}
The single-body wavefunctions depend on two length scales: the spatial separation $d$ between the two layers in the direction $\hat{z}$, and the finite layer width $w$ of each layer. In this work we assume $\phi_z(z)$ is set by an infinite square well of width $w$, 
\begin{equation}
\phi_z(z)=\sqrt{\frac{2}{w}} \sin\left(\frac{\pi z}{w}\right). 
\end{equation}
 
The Coulomb interaction in three dimensions is given by:
\begin{align}
	V_{3D}(x,y,z)=\frac{e^2}{\epsilon \ell_B}\frac{\ell_B}{\sqrt{x^2+y^2+z^2}},
\end{align}
We can then recover the Coulomb interaction part of Eq.~(\ref{ham}) by integrating out the perpendicular coordinate
\begin{multline}
	V_C^{\mu\nu}(\r)=\int\! \D z \, \D z^\prime \, |\phi_z(z)|^2 |\phi_z(z^\prime)|^2
	\\ V_{3D}(\r,z - z^\prime +  (1-\delta^\mu_\nu) d) .
\end{multline}
Throughout this work we project the Hamiltonian \eqref{ham} into the lowest Landau level,  ignoring the effects of ``Landau level mixing"  present at finite $ \frac{e^2}{\epsilon\ell_B} / \hbar \omega_c$.
In this case, it is possible to expand $V_C$ in terms of the Haldane pseudopotentials $V_\alpha$, which are the potentials felt by particles orbiting around one another in a state with relative angular momentum $\alpha$.
Later in this work we add additional $V_\alpha$ terms to $V_C$ in order to explore the neighboring phases.
In experiment, such variations of the interaction may arise due to Landau level mixing~\cite{BisharaNayakMixing, WojsMixing, RezayiSimonMixing, Papic12, SodemannMixing, PetersonMixing, SimonRezayiMixing, ZaletelMixing, PakrouskiMixing}.

Henceforth, we set the energy and length scales $\frac{e^2}{\epsilon\ell_B} = \ell_B = 1$ whenever units are omitted.

\subsection{Numerical methods}\label{sec:numerics}

We work in the Landau gauge, $(A_x,A_y)=\ell_B^{-2}(y,0)$, where the single-particle orbitals with momentum $k_x=\frac{2\pi m}{L}$ ($m \in \mathbb{Z}$) are spatially localized near $y = k_x \ell_B^2$. 
The system is fully periodic along the $x$-direction, but naturally maps to a long-range interacting 1D fermion chain along $y$-axis.
We study such chains using exact diagonalization as well as density-matrix renormalization group~\cite{ZaletelQHdmrg13, ZaletelMixing}.

For the purposes of exact diagonalization (ED), it is useful to minimize the finite-size effects by assuming the 1D chain to be periodic (i.e., the physical system is  periodic along both $x$ and $y$ directions, or equivalently it has the topology of a torus). 
Using magnetic translation symmetry reduction of the Hilbert space~\cite{Haldane-PhysRevLett.55.2095}, it is possible to study systems of about 10 electrons with pseudo-spin degree of freedom at filling $1/3+1/3$. 
The advantages of ED method are the direct access to the entire low-lying excitation spectrum, resolved ground state degeneracy, the ability to simulate complicated interactions (e.g., 3-body) that give rise to non-Abelian states, and compute overlaps between model wavefunctions and exact states.

Because of the exponential cost of ED that becomes prohibitive for systems with pseudo-spin degree of freedom, the bulk of our results are obtained via the recently developed infinite DMRG method (iDMRG)~\cite{ZaletelQHdmrg13, ZaletelMixing} that allows access to larger system sizes. 
iDMRG places the Hamiltonian on an infinitely long cylinder of circumference $L$, and employs a variational procedure to find the ground state within the variational space of matrix product states (MPS)~\cite{Fannes1992,Ostlund1995,Rommer1997}. 
MPS can only represent systems with a finite amount of entanglement $S$, which in turn is limited by the ``bond dimension" $\chi$ via $S < \log(\chi)$, while the computational resources required scale as $O(\chi^3)$. In this work we used a bond dimension $\chi \sim 5000\mbox{--}8000$.
On a cylinder, the entanglement scales with the circumference $L$, but is independent of the length of the cylinder. 
Therefore, while the complexity remains exponential in the circumference, it is constant in the length of the cylinder, which provides an advantage over ED.

\subsection{Entanglement invariants for the identification of FQH phases}
\label{sec:ent_sig}

All of the phases we study in this work are gapped,  have quantized Hall conductance $\sigma^{xy} = \frac23 (e^2/h)$, and  have no local order parameter which can be used to distinguish between them.
However, these phases do have different topological orders, and we can therefore apply a number of recent developments~\cite{Zhang2012,Cincio2013,ZaletelQHdmrg13,HHTuMomPol13} which demonstrate how the topological order of a system can be extracted from its entanglement properties.

In a topological theory, the ground state degeneracy on both the torus and infinitely long cylinder is equal to the number of anyon types.
There is a special basis for the ground state manifold, the minimally entangled basis, in which each basis state $\ket{a}$ can be identified with an anyon type $a$~\cite{KitaevPreskill, LiHaldane, Zhang2012}.

By measuring how various entanglement properties of $\ket{a}$ scale with the circumference $L$, we can measure:
	the quantum dimensions $d_a$~\cite{KitaevPreskill,LevinWen};
	the internal quantum numbers (spin, charge, etc.) of each anyon $a$;
	the ``shift" $\shift$~\cite{wenzee}, or equivalently the bulk Hall viscosity~\cite{ZaletelQHdmrg13};
	the topological spins $\theta_a = e^{2 \pi i h_a}$ and the chiral central charge $c_-$ of the edge theory~\cite{Zhang2012, ZaletelQHdmrg13, HHTuMomPol13}.
Below we provide a brief summary of these measurements in the context of FQH systems, and refer to Refs.~\onlinecite{ZaletelMixing} for a detailed discussion. 

To measure entanglement properties we divide the cylinder \emph{in orbital space} into two semi-infinite halves $L/R$ and Schmidt decompose the state as $ \ket{\Psi} = \sum_\mu \lambda_\mu \ket{\mu}_{ L} \otimes \ket{\mu}_{ R}$.
The entanglement entropy is defined as $S= - \sum_\mu \lambda_\mu^2 \log \lambda_\mu^2$.
In ground state $\ket{a}$, the entropy $S_a$ scales as \cite{KitaevPreskill,LevinWen} 
\begin{align}
	S_a = \beta L - \log\frac{\mathcal{D}}{d_a} + \mathcal{O}(e^{-L / \tilde{\xi}}),
	\label{eq:TEE}
\end{align}
where $d_a$ is the quantum dimension of anyon $a$, and $\mathcal{D}$ is the total quantum dimension of the topological phase.
The corrections are set by a length scale $\tilde{\xi}$ which need not be directly related to the physical correlation length.

To measure a \Uone charge $Q_a$ for anyon $a$, we partition the total charge operator into its components to the left / right of an entanglement cut, $\hat{Q} = \hat{Q}_L + \hat{Q}_R$. 
The left Schmidt states are eigenstates of $\hat{Q}_L$, $\hat{Q}_L \ket{\mu; a}_L \equiv Q_{\mu; a} \ket{\mu; a}_L $, where $\ket{\mu; a}_L$ are the Schmidt states of ground state $\ket{a}$ and $Q_{\mu; a} \in \mathbb{Z}$ in units where the elementary charge is 1.
The charge $Q_a$ of anyon $a$ is given by the charge polarization in the ground state, which can be expressed as an ``entanglement average'' \cite{ZaletelQHdmrg13}
\begin{align}
	e^{2 \pi i Q_a} \equiv e^{2 \pi i \sum_\mu \lambda^2_\mu Q_{\mu; a}}.
\end{align}
$Q_a$ is defined modulo 1.
In the bilayer systems with $\Uone\times\Uone$ symmetry we can apply the measurement for both layers to get two charges.

Rotating the cylinder can also be viewed as a \Uone charge, whose generator is the momentum $\hat{K}$.
Its eigenvalues $K_a$ can be combined with certain analytically calculable properties of the Landau levels to recover 
the Berry phase for an adiabatic Dehn twist (modular transformation).
Similar to the charge, the resulting phase $T_a = \exp(2\pi i M_a)$ may be computed from an entanglement average:
\begin{align}
	M_a &=  \sum_\mu \lambda^2_\mu K_{\mu; a} + \textrm{analytic terms} .
\end{align}
$M_a$ is the ``momentum polarization'', scaling as \cite{ZaletelQHdmrg13,HHTuMomPol13}
\begin{align}
	M_a &= -\frac{\nu\shift}{(4\pi\ell_B)^2}L^2 + h_a - \frac{c_-}{24} + \mathcal{O}(e^{-L / \tilde{\xi}}) \pmod{1}.
	\label{mompol}
\end{align}
Here $\shift$ is the shift, $h_a$ is the topological spin of anyon $a$, and $c_-$ is the chiral central charge of the edge.

The shift $\shift$~\cite{wenzee} is an constant mismatch between the number of flux $N_\Phi$ and electrons $N_e$ required to realize the ground state of the phase on the sphere, $N_\Phi= N_e / \nu - \shift$, and plays a particularly important role in our analysis.
For the $(330)$, $(112)$,  $\barthird$ states and the interlayer-Pfaffian (introduced in Sec.~\ref{sec:nonabelian} below) the shift takes values $\shift = 3, 1, 0, 3$ respectively (see Tab.~\ref{tab:candidates}), so distinguishes most of the phases.
Because $\shift$ in these cases is an integer and the dominant contribution to $M_a$, it converges very quickly and is far easier to measure than $h_a$, $c_-$ or $d_a$.

\begin{table}[tb]
	\begin{tabular}{c|ccccc|c}
		FQH Phase	&	\begin{tabular}{c}Ground-state\\degenacy\end{tabular}	&	$\shift$	&	\begin{tabular}{c}Spin-charge\\separation\end{tabular}	&	$c^-$
	\\\hline\hline
		(330)		&9&3&&2
	\\	(112)		&3&1&&0
	\\	\barthird	&3&0&&0
	\\	$\mathbb{Z}_4$ Read-Rezayi \cite{Read-PhysRevB.59.8084}		&15&3&&2
	\\	Interlayer-Pfaffian \cite{Ardonne02}		&9&3&\checkmark&5/2
	\\	Bonderson-Slingerland	\cite{BondersonSlingerland}	&9&4&\checkmark&5/2
	\\	Intralayer-Pfaffian \cite{Barkeshli10}		&27&3&\checkmark&3
	\\	Bilayer Fibonacci \cite{Vaezi14}		&6&?&&14/5
	\end{tabular}
	\caption{%
		Possible candidate states at $\nu=1/3+1/3$ and their observed properties.
		We call a phase ``spin-charge separated" if one can consistently assign charge/spin to the excitations, with one such excitation having neutral charge and pseudo-spin $\pm1/2$ (see Sec.~\ref{sec:nonabelian}).
	}
	\label{tab:candidates}
\end{table}

\section{Abelian Phase Diagram}
\label{sec:phase}

	In this Section we study the $\nu=1/3+1/3$ QHB system as a function of experimentally relevant parameters: interlayer separation ($d$), tunelling ($\DSAS$), and layer width ($w$). 
We determine the phase diagram using the topological characterization explained in Sec.~\ref{sec:ent_sig}, and find three different Abelian phases~\cite{HaldaneDiagram}:  decoupled $\nu = 1/3$ bilayers $(330)$ or the bilayer Laughlin phase, a bilayer-\SU2 symmetric spin-singlet hierarchy state $(112)$, and a transversely polarized particle-hole conjugate of the Laughlin state $\barthird$. 

Fig.~\ref{phase} shows the phase diagram at well width $w=0$ and cylinder circumference $L = 14$. 
Phase boundaries were determined at the points marked in black; these points were found by performing simulations in sweeps, changing either $d$ or $\DSAS$, and plotting the results. 
We find points where the correlation length and entanglement entropy have either discrete jumps or peaks, and we claim that these points are the phase transitions.
The upper panels of Figs.~\ref{Aplots}, \ref{Bplots} and \ref{Cplots} show examples of the correlation length data used to determine the locations of these transitions. The dashed lines in Fig.~\ref{phase} show the sweeps where these data were taken. 
We note that the region in the vicinity of the tentative triple point is somewhat difficult to resolve, but we have have not found any evidence for additional phases.
The three Abelian phases can be intuitively understood in the following limiting cases.

First, when $\DSAS$ is small and $d$ is large the two layers interact only weakly, and we have two decoupled Laughlin states.
Second, when $\DSAS$ is extremely large the single particle orbitals are superpositions of both layers. 
Both symmetric and antisymmetric superpositions are possible, but when $\DSAS$ is very large the antisymmetric superpositions are energetically forbidden (the energy difference between the two states is $\DSAS$), so we can view the system as a single quantum well with $\nu=2/3$,  whose ground state is the particle-hole conjugate of the Laughlin $1/3$ state, which we call the $\barthird$ state. 
This state is particularly natural at $d=0$, where the system is equivalent to a single layer with spin: the tunnelling term is a Zeeman field which spin-polarizes the system along the transverse direction.

Third, when $d=0$ and $\DSAS=0$ the system is equivalent to a single-layer system with spin that has full \SU2 symmetry.
The ground state is a $(112)$ state~\cite{Wu93, HaldaneDiagram, Moore97}. 

The attentive reader might note that, topologically, the $(112)$ and $\barthird$ phases are actually the same phase, in the sense that their $K$ matrices are related by an $\mathrm{SL}(2,\mathbb{Z})$ transformation. 
However, in the presence of rotational symmetry  these phases have a different shift $\shift$, and so they are not the same phase.
One may be concerned that in an experiment disorder will break the rotational symmetry and allow the $(112)$ and $\barthird$ state to be continuously connected, but this is in fact not the case, as this transition has been seen experimentally both in wide quantum wells~\cite{Suen94}, and in single-layer systems with spin~\cite{Eisenstein90}.

\subsection{Determination of the phases}\label{sec:determination}

\begin{figure}[ttt]
	\includegraphics[width=\linewidth]{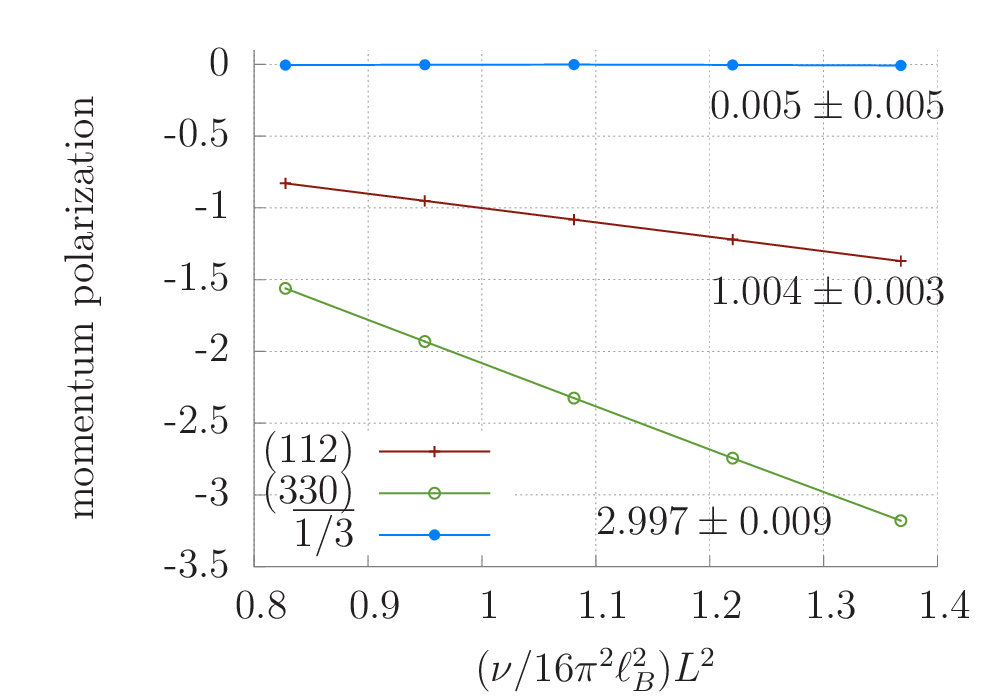}
	\caption{(Color online)
		Momentum polarization $M_a$ for the representative points from the phases in Fig.~\ref{phase}, plotted against $\frac{\nu}{(4\pi\ell_B)^2}L^2$.
		The coefficient of proportionality is the shift $\shift$, which we can read off to be $3$, $1$ and $0$ for the $(330)$, $(112)$ and $\barthird$ phase respectively, as expected.
		Data was taken at $d=1.6$, $\DSAS=0$; $d=0.2$, $\DSAS=0$; and $d=2$, $\DSAS=0.1$ for the $(330)$, $(112)$ and $\barthird$ phases, respectively. Values for the shift obtained from fitting the data are shown directly on the figure.
	}
	\label{mompolfig}
\end{figure}
We have determined the phases by using the entanglement invariants discussed in Sec.~\ref{sec:ent_sig}.
First, we measure the momentum polarization $M_a$  in order to compute the shift $\shift$, which should take the values $3$, $1$ and $0$ in the $(330)$, $(112)$ and $\barthird$ state, respectively. 
Fig.~\ref{mompolfig} shows the momentum polarization at three representative points in the phase diagram.
We plot $M_a$  as a function of $L^2$, so by Eq.~(\ref{mompol}) we should get straight lines with a slope proportional to $\shift$.
The green line $(330)$ was taken at $d=1.6$, $\DSAS=0$, giving $\shift \approx 3$;
the red line $(112)$ was taken at $d=0.2$, $\DSAS=0$, giving $\shift \approx 1$;
the blue line $\barthird$ was taken at $d=2$, $\DSAS=0.1$, giving $\shift \approx 0$.
All of these values match those predicted for the appropriate phase.

\begin{figure}[t]
	\includegraphics[width=\linewidth]{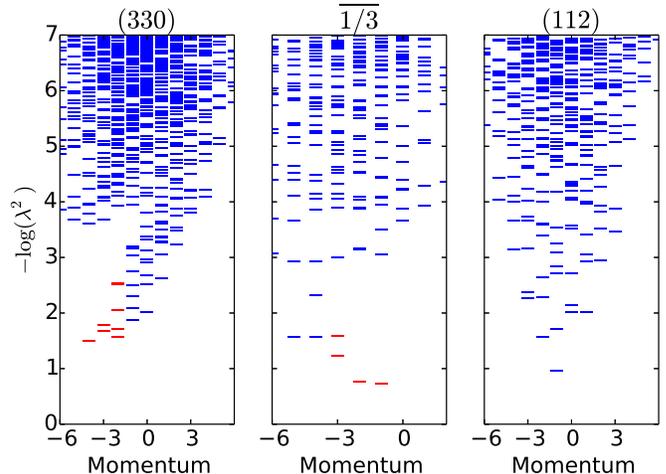}
	\caption{(Color online)
		Entanglement spectra for the phases in Fig.~\ref{phase}:
		the $(330)$ state, with counting of $1,2,5,\dots$  dispersing to the right;
		the $\barthird$ state,  with  counting $1,1,2,\dots$ dispersing to the left;
		the $(112)$ state, which has an non-chiral spectra (being a convolution of a left and right mover).
		These results are in agreement with the predicted values for these phases.
	}
	\label{spectra}
\end{figure}
Fig.~\ref{spectra} shows entanglement spectra for the same points as those shown in Fig.~\ref{mompolfig}.
The counting and chirality of the low-lying entanglement spectra are unique to each phase, and as elaborated in Fig.~\ref{spectra} we find spectra consistent with each phase.

The phase diagram in Fig.~\ref{phase} was taken using an infinite cylinder with a circumference $L=14$. 
To assess the finite size effects, we have measured the behavior of select cuts along the phase boundaries for $L=12\mbox{--}16$.
We found that the location of the $(112)\rightarrow(330)$ transition changes with system size by $d<0.02$.
The $(330) \to \barthird$ and $(112) \to \barthird$ transitions do move to smaller $\DSAS$ at larger $L$,  with a change from $L: 12 \to 16$ of about $0.003$.
While the transition may continue shifting to slightly smaller $\DSAS$ as $L$ is further increased, at large $d$ the change is small on the scale of the full phase diagram. 

At smaller $d$, the critical value of $\DSAS$ is fairly small at $L=14$ and so we may be concerned that in the thermodynamic limit it is actually zero.
We can test this at $d = 0$ by exploiting the fact that tunneling acts as a simple Zeeman field in the spin realization, so the energetics can be fully determined by the energy difference between the $(112)$ and $\barthird$ phases at $d=0$, $\DSAS=0$. 
Using the additional symmetries at this point we can perform accurate finite-size scaling to extract the energy difference in the thermodynamic limit, and we find that the transition occurs at $\DSAS\approx0.018$.
Therefore at least at small $d$, it appears that we have reached large enough sizes so that finite size effects do not change the location of the phase transition. Note that this system is formally equivalent to a $\nu=2/3$ system with spin, and our value for the energy difference matches the numerical literature for the spin-polarization transition in that system.\cite{spinpolED}

We have also assessed the sensitivity to layer width $w$ for select cuts through the phase boundary.
In the upper panels of Figs.~\ref{Bplots} and \ref{Cplots}, we used dashed lines to show the correlation lengths at finite widths.
We see that a finite layer width shifts the location of the $(112):(330)$ transition to larger $d$, while the $(330):\barthird$ transition is shifted to smaller $\DSAS$. 
At $w = 1$ the boundaries have changed by about 10\% compared to $w = 0$, so we don't expect any qualitative differences in the phase diagram.

Naturally there are many differences between the system we are studying numerically and those which are studied in experiments.
 In addition to the finite-size effects and our simplified treatment of layer width, we also neglect other factors including Landau level mixing and  disorder. 
 One can therefore ask how relevant our data is to experiments, particularly as to the quantitative locations of the phase transitions shown in Fig.~\ref{phase}.
 One way to address this is to compare to the experimental data which already exists.
Ref.~\onlinecite{Hirayama2002b} studied the $(330):\barthird$ transition and found it at approximately $d=2$, $\DSAS=0.1$.
The location of their observed transition is shown in Fig.~\ref{phase}. 
We obtain $\DSAS \approx 0.07$, and this gives us reason to believe that our data can be used as a guideline for future experiments. 

\subsection{Order of the transitions}\label{sec:order}

\begin{figure}[ttt]
	\includegraphics[width=\linewidth]{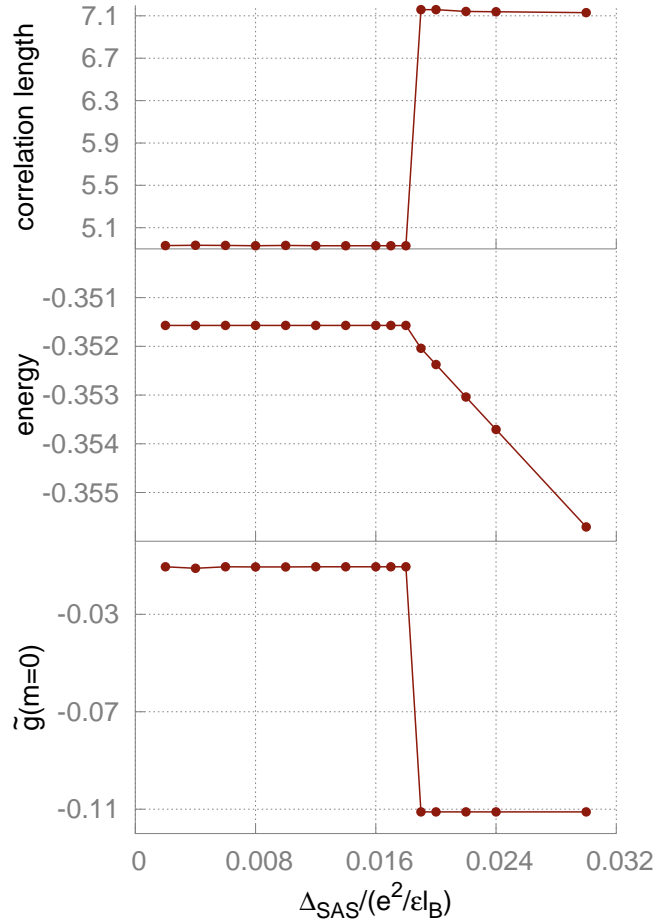}
	\caption{\label{Aplots} (Color online) Data as a function of tunneling strength, crossing the $(112):\barthird$ transition. The correlation length is flat except very close to the transition, where it is discontinuous. There is also a kink in the energy and in $\corrM$. This is all consistent with a first-order transition. }
\end{figure}

\begin{figure}[ttt]
	\includegraphics[width=\linewidth]{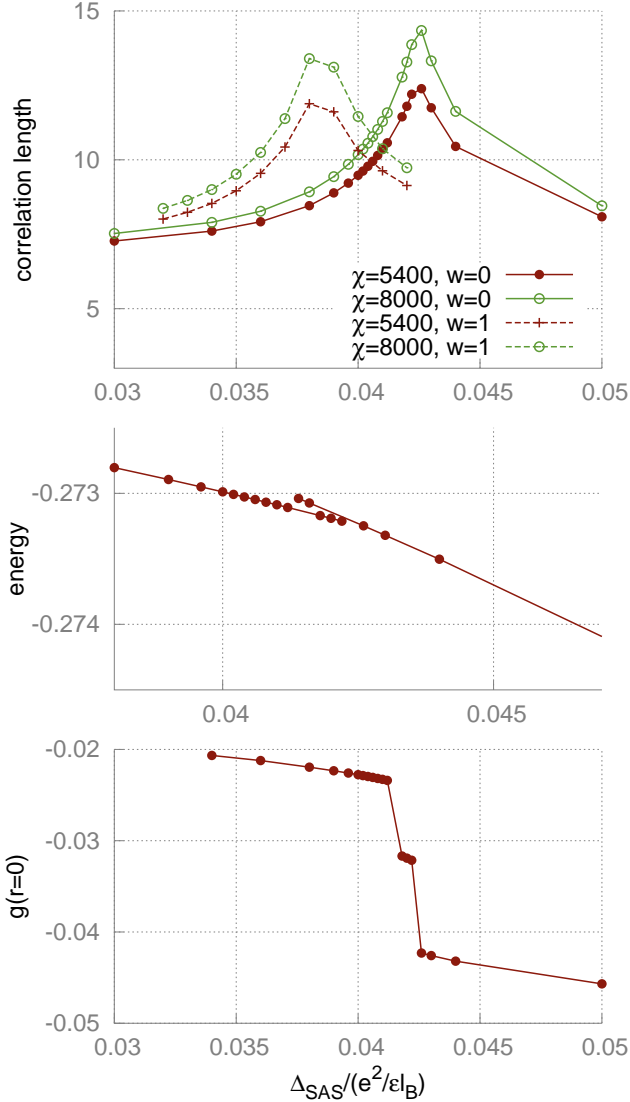}
	\caption{\label{Bplots} (Color online) Data as a function of tunnelling strength, crossing the $(330):\barthird$ transition. The correlation length has a peak near the transition, but this is consistent with both a first and second order transition. The middle panel shows the energy for both the $(330)$ and $\barthird$ phases (see text), and as these lines are not parallel the systems energy has a kink. There is also a jump in $\corrR(r=0)$, consistent with a first-order transition. }
\end{figure}

The large system sizes accessible to our DMRG simulations allow us to assess the nature of the various phase transitions in Fig.~\ref{phase}. 
We find strong evidence that the $(330):\barthird$ and $ (112):\barthird$ transitions are first order. The $(330):(112)$ transition appears to be very weakly first order, though we cannot definitely rule out a continuous transition.
To determine the order of the transition we check for discontinuities in $\partial_g E$, where $g = \DSAS, d$ tunes across the transition, as well as for divergences in the correlation length and discontinuities in local observables.

\begin{figure}[ttt]
	\includegraphics[width=\linewidth]{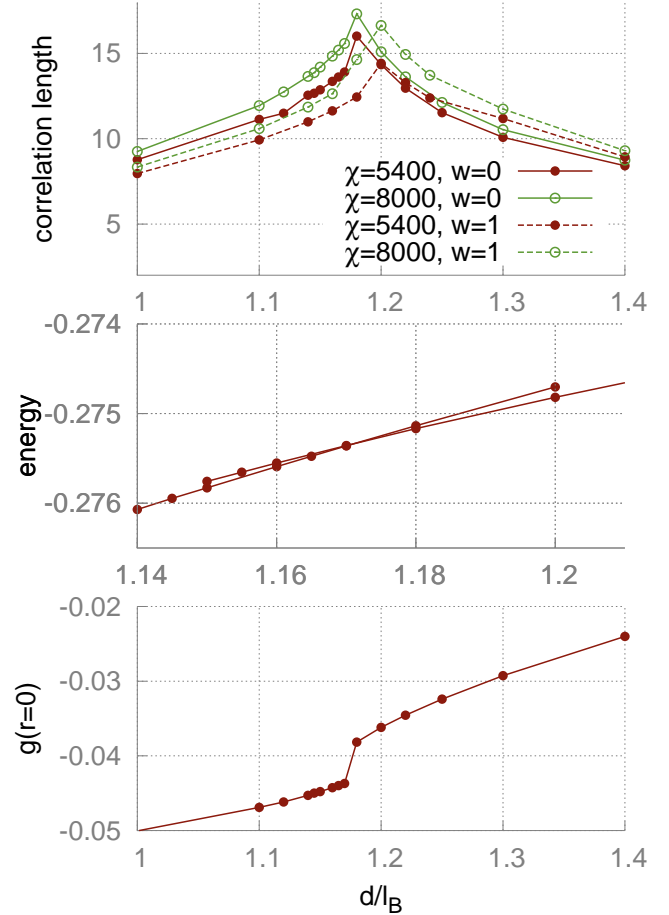}
	\caption{\label{Cplots} (Color online) Data as a function of interlayer separation, crossing the $(330):(112)$ transition. The correlation length has a peak, while the energy has a kink and the $\corrR(r=0)$ are jumps across the transition. This is indicative of a first order transition, though the transition is weaker compared to the others in the phase diagram. }
\end{figure}

The upper panel of Fig.~\ref{Aplots} shows the $(112):\barthird$ transition, at which the correlation length  jumps discontinuously while remaining finite, indicating a strongly first-order transition. 
In the upper panels of Figs.~\ref{Bplots} and \ref{Cplots} we show correlation lengths for $(330):\barthird$ and $(330):(112)$ transitions.
The correlation length peaks as the transition is approached, suggesting either a continuous or weakly first order transition.
A  continuous transition would be gapless, generating a large amount of entanglement which cannot be efficiently represented by an MPS; finite $\chi$ effects then cutoff the divergent $\xi$.
Consequently we would expect  a strong dependence of $\xi$ on the MPS bond dimension $\chi$.
The different colored lines in the figure correspond to increasing $\chi$, and we  see that $\xi$ increases with $\chi$, which could be consistent with a continuous transition. 
However, a similar effect could be seen at a weakly first-order transition if $\chi$ is not large enough to capture the state.
Therefore we need other ways to determine the order of these transitions.
 
Another approach is to look at behaviour of the energy at the transition point. For a first-order transition, we expect a kink in the energy, while for a continuous transition we expect the energy to vary smoothly. The middle panels of Figs.~\ref{Aplots}, \ref{Bplots} and \ref{Cplots} show the energies near these transitions. 
The first order $(112):\barthird$ transition has a clear kink in the energy.
The $(330):\barthird$ transition also appears of have a kink. 
The system also exhibits hysteresis for both the $(112):\barthird$ and $(330):\barthird$ transitions: if we initialize the system in the $\barthird$ phase it will stay in that phase even if $\DSAS$ is below its critical value. 
This is of course expected in a first order transition, and in the middle plot of Fig.~\ref{Bplots} we plot two separate lines, which are the energy of the $(330)$ and $\barthird$ phases (the actual energy of the system is whichever of these energies is lower). 
We can see that these lines are not parallel, which clearly shows that there is a kink in the system's energy and therefore the transition is first order.
At the $(330):(112)$ transition we find a very weak kink, so we tentatively conclude all three transitions are first order.

It is also useful to look at the behavior of local correlations, such as the real space density-density correlation between electrons in different layers:
\begin{align}
	\corrR(\r)=\braket{ n^\up(\r)n^\dn(0) } - \braket{n^\up(\r)} \braket{n^\dn(0)},
\end{align}
where $n^\mu(\r)$ was defined in Eq.~(\ref{density}).
In the $(330)$ phase, the layers are uncorrelated, and this quantity should be approximately zero. In the other phases, at small $\r$ the electrons repel and so $\corrR(\r)$ should be negative. We can also look at the same correlation function in orbital space instead of real space:
\begin{align}\begin{split}
	\corrM(m) &= \braket{ n^\up_m n^\dn_0 } - \braket{n^\up_m}\braket{n^\dn_0}, \\
	n^\mu_m &\equiv {c_m^\mu}^\dagger c^\mu_m.
\end{split}\end{align}
For $m=0$, this quantity will be negative in the $\barthird$ phase, but it will be small in the other phases. 
When the above quantities have different values on either side of a phase transition, we expect them to jump discontinuously for a first-order transition and to vary continuously for a second-order transition.

We plot these quantities in the bottom panels of Figs.~\ref{Aplots}, \ref{Bplots} and \ref{Cplots}, and see discrete jumps in all cases.
Based on the results of this section we can claim that all the transitions in the diagram are first order, with the strongest first order transition being the $(112):\barthird$ transition. The $(330):(112)$ transition has only a slight kink in the energy and the jump in $g(r)$ is smaller than the other transitions, so this is the weakest first order transition in the diagram.

In Ref.~\onlinecite{Hirayama2004}, four experimental sweeps in our phase diagram were performed. Two of these sweeps had small $\DSAS$, and had $d\approx1.4-2.8$. These sweeps found a $\nu=2/3$ state which we take to be the $(330)$ state at large $d$, but below $d\approx1.8$ they find no QH state. 
We believe that this is because their experiments were taken at layer width $w/\ell_B\approx2$, which would move the $(330):(112)$ transition to larger $d$, putting it near where they observe the vanishing QH state. 
Furthermore, we have found that the $(330):(112)$ transition is weakly first-order, implying that at the transition there is a small energy gap. 
We surmise that the quantum Hall state is not observed in experiment because the gap is very small near the transition, and so the transition point is being smeared by finite temperature and disorder effects.

\subsection{Spin polarization}\label{sec:polarization}

In addition to the bilayer degree of freedom electrons carry spin, resulting in a four-component system.
Thus far we have assumed the spin is polarized by the external magnetic field, an assumption we can test with our simulations. 

The spin-polarized $\barthird$ phase at $d=0, w = 0$ and large $\DSAS$ is essentially a one-component system with filling $2/3$, while the competing spin-unpolarized state is a two component (spin) system with each component having filling $1/3$. 
The spin-unpolarized case has a lower Coulomb energy proportional to $\ell_B^{-1} \propto B^{1/2}$ (this is why we find $(112)$ in the equivalent bilayer problem), while the spin-polarized state  gains a Zeeman energy  proportional to the applied field $B$. 
For systems at fixed $\nu=2/3$, for a small perpendicular magnetic field (and proportionally small density), the system will be in a spin-unpolarized state, while for large magnetic field (and density) the system will spin polarize.
The spin base case been studied both numerically \cite{spinpolED} and experimentally \cite{Eisenstein90}, but the results to not agree, with the numerics predicting a critical magnetic field of $\approx \unit[11]{T}$ and experiments measuring $\approx \unit[3]{T}$. 
It has been proposed that the difference between these values is due to the finite layer width of the samples \cite{Eisenstein90}. 
We are in a position to confirm this, and indeed we find that increasing the layer width does decrease the critical magnetic field, with a layer width of $\approx5$ magnetic lengths being sufficient to bring experiment and simulation into agreement. 
Thus, in context to the bilayer set up, whether the $\barthird$ state is completely spin-polarized will depend on the bilayer separation $(d)$ and the strength of the magnetic field.


For the bilayer-$(112)$ point at $d = 0, w = 0$  we compute the energy of an \SU4 symmetric four-component system (bilayer + spin) with each component having filling $1/6$.
The resulting state is gapless, which means that our DMRG performs poorly and we can only obtain a rough estimate for the energy.
However, it appears that the magnetic field required to spin-polarize the system is approximately an order of magnitude less than that required to polarize the $\barthird$ phase, so this phase should be spin-polarized even at small magnetic fields.

In the large-$d$ $(330)$ phase, the problem reduces to decoupled layers, and it is well known that $\nu = 1/3$ system spin-spin polarizes, so we expect this will remain true for all $d$ into the $(112)$ phase. 

Also note that experimental studies \cite{Hirayama2002b,Hirayama2004} on this system have observed a spin-polarized system at all the tunnelling strengths and interlayer separations accessed, for magnetic fields $B \approx 4\mbox{--}\unit[11]{T}$.



\section{Non-Abelian phase}
\label{sec:nonabelian}

\begin{figure}[ttt]
	\includegraphics[width=\linewidth]{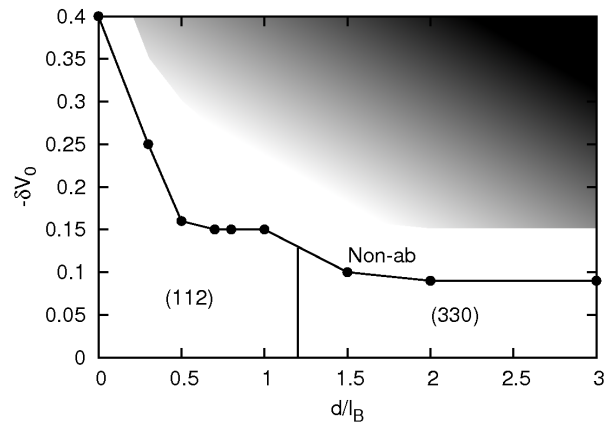}
	\caption{Phase diagram as a function of interlayer separation $d$ and the modification of the Haldane potential $\delta V_0$. We find that as $-\delta V_0$ is increased, a new phase appears which we believe is a bilayer-spin charge separated non-Abelian phase. Data is taken with zero tunneling $\DSAS = 0$ and layer width $w=0$.}
	\label{V0phase}
\end{figure}

In addition to the Abelian phases shown in Fig.~\ref{phase}, a number of non-Abelian candidates have been proposed to appear in the $1/3+1/3$ system.
These include the $\mathbb{Z}_4$ Read-Rezayi state~\cite{Read-PhysRevB.59.8084}, the ``interlayer-Pfaffian" (iPf)~\cite{Ardonne02} and ``intralayer-Pfaffian" states~\cite{Barkeshli10}, and the bilayer Fibonacci state~\cite{Vaezi14}.
While we find no signature of these non-Abelian phases when restricting to the lowest Landau level and tuning the parameters $d$, $w$, and $\DSAS$, experimental samples certainly contain further tuning parameters we have neglected.
To account for those, we have further perturbed the model with Haldane pseudopotentials $V_0$ and $V_1$.
Remarkably, we find that a modification of the interlayer interaction, either through an attractive hard core $-\delta V_0$ or repulsive hollow-core $\delta V_1$, is sufficient to drive the system into a non-Abelian phase over a range of layer separations $d$. 
In Fig.~\ref{V0phase} we show the phase diagram at fixed $\DSAS = 0$, $w = 0$, as we scan $d$ and the interlayer perturbation $-\delta V_0$.
We find that for all interlayer separations $d$ it is possible to reduce $V_0$ enough to reach a new phase consistent with the interlayer-Pfaffian (iPf) state, the evidence for which we present in this section.

\begin{figure}
	\includegraphics[width=\linewidth]{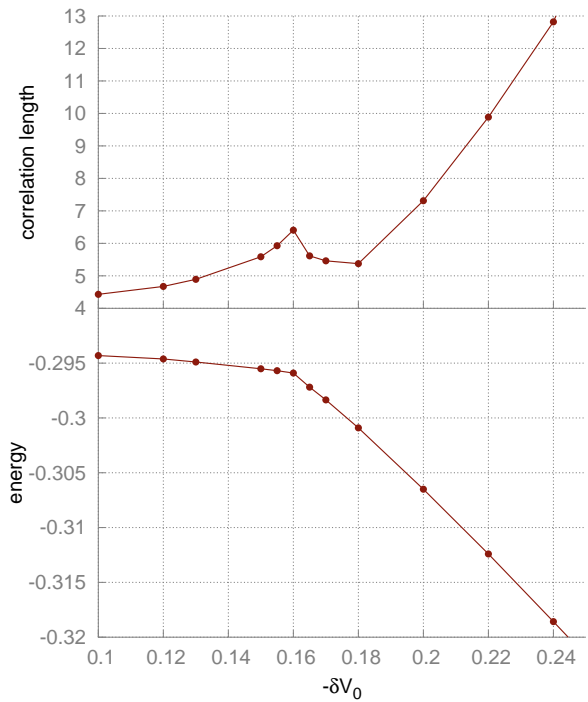}
	\caption{(Color online) 
		Correlation length and energy for the spin-charge separated state as a function of $\delta V_0$ for $d=0.5$, showing a clear first-order transition at $\delta V_0 \approx 0.16$.
		Note that correlation length increases rapidly as $V_0$ is made further negative.}
	\label{V0sweep} 
\end{figure}

Fig.~\ref{V0sweep} shows a plot of correlation length and energy as a function of $\delta V_0$ for $d=0.5$. 
There is clearly a peak in the correlation length and a kink in the energy at $\delta V_0 \approx 0.16$, indicative of a first-order phase transition.
The other points in Fig.~\ref{V0phase} were determined from similar data. 
As $-\delta V_0$ is  increased much further, we see that the correlation length continuously increases, and eventually the iDMRG becomes unstable [shaded area in Fig.~\ref{V0phase}].
Based on small systems studied by ED, in this regime we expect a strongly-paired phase where electrons form tightly bound pairs in real space~\cite{Halperin83, ReadGreen}.
Upon even further increase of $-\delta V_0$ [not shown in Fig.~\ref{V0phase}], using ED we find symmetry-broken, CDW and clustered phases~\cite{KunYang14}.

In the new intermediate $\delta V_0$  phase the iDMRG finds two nearly-degenerate ground states which we label $\ket{\Omega^1}$ and $\ket{\Omega^2}$.
These states in fact triple the unit cell along the cylinder, so by translating $\ket{\Omega^1}, \ket{\Omega^2}$ we know there are \emph{at least} six ground states in total.
This must be understood as a lower bound on the degeneracy, as there is no general way to guarantee  iDMRG finds all possible ground states.

Our evidence for identifying the novel phase with the iPf is five-fold.

\begin{enumerate}
\setlength\parskip{0mm}
\setlength\parsep{0mm}
\setlength\itemsep{2mm}
	\item	The shift  is $\shift=3$, as determined by the momentum polarization.
	\item	From the ground state $ \ket{\Omega^2}$ we deduce there is a anyonic excitation that carries pseudo-spin $\pm\frac12$ yet is charge neutral. Hence the phase is ``spin-charge separated", and we call this excitation the spinon.
	\item	The spinon excitation is  non-Abelian, with quantum dimension $d_{\Omega^2} \approx 1.4$ consistent with the iPf but not the intralayer-Pfaffian.
	\item	The momentum polarization of the two ground states differ by $h_{\Omega^2} - h_{\Omega^1} \approx -0.21$, which corresponds to the difference in the topological spins of the associated anyons.
	\item	The ground states exhibit a purely chiral entanglement spectra with counting that varies with charge sector.
\end{enumerate}
A summary of the possible candidates is listed in Tab.~\ref{tab:candidates}
These observations eliminate all other known candidates for the $1/3+1/3$ system.
In the following Sections, we give a brief description of the iPf phase (Sec.~\ref{sec:iPf}), compute overlaps against the model wavefunction using ED (Sec.~\ref{sec:ed_overlap}), and present evidence
for  spin-charge separation (Sec.~\ref{sec:scsep}) 
and  non-Abelian statistics (Sec.~\ref{sec:nonabelian_sig}).

\subsection{The interlayer-Pfaffian state}\label{sec:iPf}
The iPf phase was first introduced and extensively discussed in Ref.~\onlinecite{Ardonne02}, and coined the interlayer-Pfaffian in Ref.~\onlinecite{Barkeshli10}.
Similar to the Moore-Read phase relevant at $\nu = 5/2$, the interlayer-Pfaffian has non-Abelian Ising anyon excitations, which behave like unpaired Majorana zero modes.
But the iPf phase is even more interesting than the Moore-Read phase as it is ``spin-charge separated".
Here we treat the two layers as an effective spin system and label them as $\up$ and $\dn$.
The total charge is the sum $Q = Q^\up + Q^\dn$ while the ``pseudo-spin" is the difference $S^z = \frac{1}{2} (Q^\up -  Q^\dn)$. 
The \emph{local} excitations are built up from neutral excitons and electrons.
The neutral bilayer-excitons  have $Q=0$ and carry integral $S^z =  0, \pm 1, \pm 2, \dots$, while the $Q = 1$ electrons carry $S^z = \pm \frac{1}{2}$. 
Thus local excitations obey the relation $Q \equiv 2S^z \pmod2$, ``locking'' spin and charge together.
In the iPf phase the electron can fractionalize into a neutral non-Abelian ``spinon'' carrying $Q = 0, S^z = \frac{1}{2}$ and three non-Abelian ``chargons'' carrying $Q = \frac{1}{3}, S^z = 0$.
Thus when including fractional excitations there are no constraints between charge and spin.

A representative (model) wavefunction for the iPf phase is given by~\cite{Ardonne02}
\begin{align}
	\Psi( \{z\}, \{w\}) &= \operatorname{Pf}\left( \frac{1}{x_i-x_j} \right) \Psi_{221}(\{z\},\{w\}).		\label{ipfwf}
\end{align}
Here $\{z\}$ and $\{w\}$ denote complex 2D coordinates of electrons in two layers, while $\{x\}=\{z,w\}$ stands for coordinates of all electrons, regardless of their layer index.
The (221) state is defined as
\begin{align}
	\Psi_{221} &= \prod_{a<b} (z_a - z_b)^2 \prod_{a<b} (w_a - w_b)^2  \prod_{a,b} (z_a - w_b)	\notag
	\\	&\quad \times e^{ -\frac14 \sum_a |z_a|^2 } e^{ -\frac14 \sum_a |w_a|^2 } .
\end{align}

There are nine anyon types in the iPf phase, which break up into three sets of three.
Three of these anyons are over-all charge neutral and form the Ising theory: the trivial sector $\mathds{1}$, a neutral fermion $\psi$ which carries fermion parity but no charge, and the non-Abelian spinon excitation $\spinon$, which  carries pseudo-spin $S^z = \pm\frac12$ but no charge
	\footnote{Technically the list of anyons are sets of excitations, rather than a single excitation.  What we've described here are the properties of a representative excitation within each set, with all other excitations related by the addition/removal of electrons.}.
In addition, threading $2 \pi$ flux quanta induces a charge $Q = \frac{1}{3} +  \frac{1}{3}$ Abelian anyon we denote by $\Phi$.
The fusion rules are
\begin{align}
	\spinon \times \psi &= \psi, & \spinon \times \spinon &= \mathds{1} + \psi, & \Phi^3 &= \mathds{1}.
\end{align}
By combining fluxes $\Phi$ with the Ising sector, we obtain the nine anyon types:
\begin{align}
	\renewcommand{\arraystretch}{1.3}
	\begin{array}{cc|ccc}
		&&	\multicolumn{3}{c}{\textrm{charge }Q}
	\\	&&	0	&	\frac23	&	\frac43
	\\\hline
		\multirow{4}{*}{\begin{turn}{90}spin $S^z$\end{turn}}&0&	\mathds{1}	&	\Phi	&	\Phi^2
	\\	&0&	\psi	&	\psi\Phi	&	\psi\Phi^2
	\\	&\frac12&	\spinon	&	\spinon\Phi	&	\spinon\Phi^2
	\end{array}
\end{align}

\subsection{Exact-diagonalization overlaps}
\label{sec:ed_overlap}
Corresponding to the nine anyon types we should obtain nine degenerate ground states on the torus or an infinite cylinder.
Using the 3-body parent Hamiltonian~\cite{lankvelt, CHLee15} for the model wavefunction in Eq.~\eqref{ipfwf}, we have verified this is indeed the case on the torus.
By performing exact diagonalization of this Hamiltonian, we find three ground states with zero momentum, each being 3-fold degenerate due to center-of-mass translations, which yields nine ground states in total.
In the ``thin-torus" limit~\cite{Seidel, Bergholtz}, these ground states reduce to the correct Tao-Thouless states, expected from Eq.~\eqref{ipfwf}.


In small systems accessible by ED, the overlap with iPf model wavefunction becomes large in the novel phase identified in Fig.~\ref{V0phase}.
For small systems up to 10 particles, we can obtain the complete set of exact ground states on the torus corresponding to Eq.~\ref{ipfwf}, and overlap those with the same number of lowest states of the Coulomb interaction (possibly with some short-range pseudopotenials added).
This defines an overlap matrix. The sum of singular values of the overlap matrix can serve as a rough indicator if the system is in the iPf phase or not. For example, singular values close to zero would indicate the system being far from the iPf phase. In a finite system, singular values that can be considered ``non-zero" are those larger than $1/\sqrt{\operatorname{dim}\cal{H}}$, where $\operatorname{dim}\cal{H}$ is the dimension of the Hilbert space. Note that because of the invariance under the center-of-mass translation, it is sufficient to restrict only to the three ground states with momentum equal to zero, i.e., we obtain a $3\times3$ overlap matrix.

\begin{figure}[tb]
	\includegraphics[width=\linewidth]{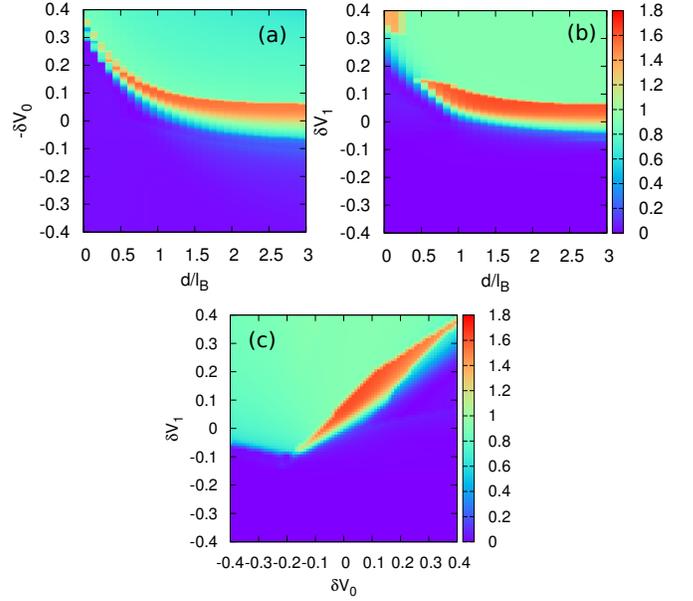}
	\caption{%
		(Color online)
		Overlap between the iPf state and the ground state of Coulomb interaction with modified short-range pseudopotentials from ED.
		The system contains 8 electrons and 12 flux quanta, on a torus with a hexagonal unit cell. The color scale indicates the sum of singular values of the $3\times3$ overlap matrix defined in the main text.
		(a) The interaction is varied by changing $d/\ell_B$ and adding $\delta V_0$ pseudopotential.
		(b) The interaction is varied by changing $d/\ell_B$ and adding $\delta V_1$ pseudopotential. The same amount of $\delta V_1$ is added to both intralayer and interlayer Coulomb. 
		(c) The effect of varying both $V_0$ and $V_1$ at fixed bilayer distance $d=1.5\ell_B$. Note that the iPf phase is located in the narrow red strip, and can be stabilized by either the reduction in $V_0$ (a), the increase in $V_1$ (b), or increase of both $\delta V_0$ and $\delta V_1$ (c).		
	}
	\label{overlap}
\end{figure}
Fig.~\ref{overlap} summarizes the effect of varying short-range $V_0$ and $V_1$ components of the Coulomb interactions inferred from the overlap of the ground state (obtained by ED) and the model wavefunction, Eq.~(\ref{ipfwf}). 
We plot the sum of singular values of the overlap matrix between the exact ground state of the Coulomb interaction (with modified short-range components) and the iPf state. In Fig.~\ref{overlap}(a),(b) we vary the bilayer distance $d$ and add $V_0$ (a) or $V_1$ pseudopotential (b) to the Coulomb interaction. The system contains 8 electrons and 12 flux quanta on a torus with a hexagonal unit cell.
We first note that the largest value of the overlap occurs in the narrow red strip, corresponding to intermediate values of $d$ and the reduction of $V_0$ or, conversely, the increase of $V_1$. The non-zero overlap in this region suggests that the system is in the iPf phase.
The ED result in Fig.~\ref{overlap}(a) can be directly compared with the phase diagram obtained by DMRG in Fig.~\ref{V0phase}.
We note that the variation $\delta V_1$ in Fig.~\ref{overlap}(b) assumes adding the same amount of $\delta V_1$ to both intralayer and interlayer Coulomb pseudopotential. Another possibility is to add $\delta V_1$ to interlayer Coulomb only. This yields a qualitatively similar result to Fig.~\ref{overlap}(b) but with somewhat stronger finite-size effects.

Finally, in Fig.~\ref{overlap}(c) we consider a combined effect of simultaneously varying $V_0$ and $V_1$. The starting point is Coulomb interaction at fixed bilayer distance $d=1.5$ in the (330) phase. In this case we find the iPf phase to be stabilized for positive $\delta V_0$ as well as positive $\delta V_1$. Note that the largest overlap (i.e., sum of singular values of the overlap matrix) is roughly the same  in all cases shown in Fig.~\ref{overlap}. Although the magnitude of the overlap with the iPf is significant, it is relatively moderate (at maximum 1.8 compared to the ``perfect" value of 3). The reason for this is the difficulty in fully resolving the complete set of iPf ground states in small finite systems. For example, finding only two out of three ground states will significantly reduce the overlaps in Fig.~\ref{overlap}. This is responsible for small overlaps in at least part of the green region in Fig.~\ref{overlap}, and leads to a somewhat narrower iPf phase compared to the DMRG result in Fig.~\ref{V0phase}.

\subsection{Spin-charge separation}\label{sec:scsep}

\begin{figure}[tb]
	\includegraphics[width=\linewidth]{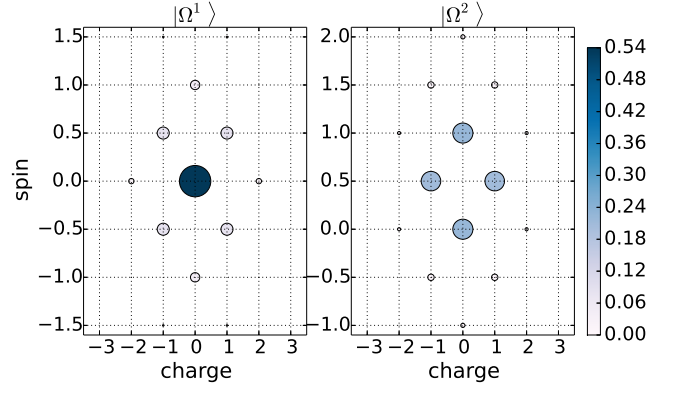}
	\caption{(Color online)
		Entanglement of for spin-charge separation in the non-Abelian phase.
		We plot the probability $P_a(Q_L, S^z_L )$ for charges $Q_L, S^z_L $ to fluctuate to the left of the cut in ground states $a =\Omega_1 / \Omega_2$. 
		The center of this distribution gives the charge and spin of the anyon associated with the ground state. 
		We see that ground state $\Omega_1$ corresponds to a quasiparticle with $S^z=0$ and $Q = 0$, consistent with either the $\mathds{1}$ or  $\psi$  sector; in the other ground state there is a quasiparticle with $S^z=1/2$ and $Q = 0$, consistent with the $\spinon$ sector.
	}
	\label{charge}
\end{figure}

We now demonstrate how we can extract the charges $(Q,S^z)$ of an anyon $a$ from entanglement spectrum of its associated ground state $\ket{a}$.
Partition  the cylinder with a cut along the circumference into ``left'' and ``right'' semi-infinite halves.
Each left Schmidt state $\ket{\beta; a}$ of the MES $\ket{a}$ has quantum numbers $Q^{\up/\dn}_{\beta; a}$.
By coarse graining the Schmidt spectrum $\lambda_{\beta; a}$ over quantum-number sectors, we can look at the probability distribution $P_a$ for charge $Q_L$ or spin $S_L^z$ to fluctuate to the left of the cut:
\begin{subequations}\begin{align}
	1 &=  \sum_{Q_L , S^z_L }  P_a(Q_L, S^z_L ) ,\\
	\langle \hat{Q}_L \rangle_a &= \sum_{Q_L, S^z_L }  P_a(Q_L, S^z_L ) Q_L, \\
	\langle \hat{S}^z_L \rangle_a &= \sum_{Q_L , S^z_L }    P_a(Q_L , S^z_L) S^z_L .
\end{align}\end{subequations}
The first equation expresses normalization.
The ``entanglement averages" in the second and third equation determine the charge and pseudo-spin of the anyon $a$ (modulo local excitations).
In Fig.~\ref{charge}, we have plotted this probability distribution in the spin-charge plane for the states $\ket{\Omega^1}$, $\ket{\Omega^2}$.
Intuitively the $\ket{\Omega^1}$ has a probability distribution associated with a completely neutral object, plus some number of electrons; in contrast $\ket{\Omega^2}$ has a probability distribution associated with a $Q = 0$, $S^z = \pm \frac{1}{2}$ object, plus some number of electrons.
The anyon associated with the latter ground state is what we identify with the spinon ($\spinon$).

Our interpretation can be made rigorous by viewing the cylinder wavefunction  as a 1D fermion chain and appealing to the theory of 1D symmetry-protected topological (SPT) phases.
The internal symmetry group of the bilayer is $G = (\Uone \times \Uone) \rtimes \mathbb{Z}_2$, coming from particle conservation in each layer and the interchange of the two layers. 
As discussed in Sec.~\ref{sec:ent_sig},  any global symmetry group $G$ can be restricted to the left half of the system in order to determine how it acts on left Schmidt states.
In a 1D symmetry-protected topological (SPT) phases, the symmetries $G$ may be represented \emph{projectively} on the Schmidt states.
The classification of 1D-SPT phases is given by the distinct possible projective representations, which are in turned classified by the second group-cohomology classes~\cite{FidkowskiKitaev2011}:
\begin{align}
	\textrm{1D $G$-symm. phases} \;\leftrightarrow\; \textrm{elements of } H^2 \big( G, \Uone \big).
\end{align}
For our symmetry the group cohomology is
\begin{align}
	H^2 \big( \Uone^2 \rtimes \mathbb{Z}_2 , \Uone \big) = \mathbb{Z}_2
\end{align}
meaning there are two possible SPT phases.
We identify the two phases with ``integral" (trivial) and ``half-integral" (non-trivial) spin. 
On an open chain, the non-trivial SPT phase has a protected two-fold degenerate edge state that transforms projectively under $G$.

The claim is that the MES $\ket{\spinon}$ has non-trivial 1D-SPT order under $G$, while $\ket{\mathds{1}}$ and $\ket{\psi}$ are trivial. 
Now suppose we create a domain wall between the $\ket{\mathds{1}}$ and $\ket{\spinon}$ topological sectors. 
The domain wall must contain an anyon excitation $\spinon$. 
However, in the 1D picture we have formed a domain wall between  trivial and non-trivial 1D-SPT phases, which \emph{must}  have an emergent edge excitation.
For our symmetry group $G$, the edge is two-fold degenerate, corresponding to the internal spin index $S^z = \pm \frac{1}{2}$ of the spinon $\spinon$ trapped there.

Referring back to Fig.~\ref{charge}, we see that $\ket{\Omega^2}$ has a 2-fold degenerate probability distribution, which is a tell-tale signature of a 1D-SPT phase.
One can explicitly check that $\ket{\Omega^2}$ is a non-trivial SPT phase under $G$, which is why we identify $\spinon \leftrightarrow \Omega^2$.

In summary, we have shown the state has an excitation with quantum numbers $Q = 0, S^z = \pm \tfrac{1}{2}$, which rules out the (330),  $\mathbb{Z}_4$ Read-Rezayi, and  Fibonacci phases.
In light of this data, we find that $\ket{\Omega^2}$ is consistent with $\ket{\spinon}$, while $\ket{\Omega^1}$ is consistent with either $\ket{\mathds{1}}$ or $\ket{\psi}$.
The absence of either the $\mathds{1}$-family or $\psi$-family from our numerics is not terribly troubling, as $\mathds{1}$ and $\psi$ have no symmetry properties which distinguish them; even a slight energetic splitting of the topological degeneracy may consistently bias the iDMRG towards the latter.

The intralayer-Pfaffian also has a $Q = 0$, $S^z = \pm \tfrac{1}{2}$ spinon, but it can be distinguished from the spinon of the interlayer-Pfaffian by its quantum dimension.

\subsection{Non-Abelian signatures}\label{sec:nonabelian_sig}

In order to directly confirm the non-Abelian nature of the novel phase, we measure the quantum dimension of the spinon.
In the iPf phase, the spinon has quantum dimension $d_{\spinon} = \sqrt{2}$. In contrast, the intralayer-Pfaffian phase has two kinds of $\spinon$ excitations, each of which lives in only the top or bottom layers.
In this phase the observed quasiparticle with spin-charge separation is a product of a spinon in each layer, and it therefore has quantum dimension $d=2$.
Our measurements of the quantum dimension therefore allow us to rule out the intralayer-Pfaffian.

\begin{figure}[tb]
	\includegraphics[angle=-90,width=\linewidth]{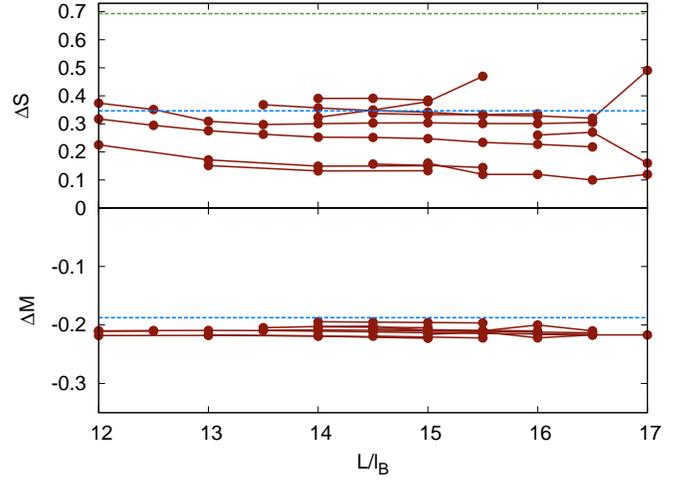}
	\caption{(Color online)
		Differences in entanglement entropy and momentum polarization for the two degenerate states as a function of circumference.
		Data was taken at a variety of different interlayer separations, $\delta V_0$ and $\delta V_1$.
		The blue dashed lines show the expected values for the iPf phase, at $S_\spinon - S_\psi = \log\sqrt{2}$ and $h_\spinon-h_\psi = -\frac{3}{16}$.
	}
	\label{deltas}
\end{figure}

To make this measurement, we compute the difference in the entanglement entropy between $\ket{\Omega^1}$ and $\ket{\Omega^2}$.
From Eq.~\eqref{eq:TEE}~\cite{KitaevPreskill,LevinWen},
\begin{align}
	S_{\Omega^2}(L) - S_{\Omega^1}(L) = \log(d_{\Omega^2}/d_{\Omega^1}) + \mathcal{O}(e^{-L/\tilde{\xi}})
\end{align}
from which we obtain the ratio of quantum dimensions $d_{\Omega_2}/d_{\Omega_1}$.
Assuming $\ket{\Omega^1}$ corresponds to a Abelian anyon ($d_{\Omega^1}=1$), and provided the finite-size effects are small enough, we extract the quantum dimension of the spin-charge separated anyon.
In Fig.~\ref{deltas} we show the results of this subtraction, for $L = 12\mbox{--}17$, for several different combinations of $d$, $\delta V_0$ and $\delta V_1$.
Finite-size and finite-$\chi$ effects introduce significant systematic errors into our calculation of this quantity, leading to results for $\Delta S = S_{\Omega^2} - S_{\Omega^1}$ which vary from $0.1-0.5$ for different measurements. Though this prevents us from determining the quantum dimension precisely, we can still say that our results are consistent with $d_{\spinon} = \sqrt{2}$ (as shown by the blue dashed line in Fig.~\ref{deltas}), and inconsistent with the intralayer Pfaffian value $d=2$ (as shown by the green dashed line), and the abelian value of $d=1$. 

Furthermore, we can use a similar subtraction scheme to extract the (relative) topological spin of the spinon ($\Omega^2$) compared to the netrual ($\Omega^1$) via the momentum polarization [Eq.~\eqref{mompol}].
Taking the difference of the momentum polarizations of the ground states
\begin{align}
	M_{\Omega^2} - M_{\Omega^1} = h_{\Omega^2} - h_{\Omega^1} + \mathcal{O}(e^{-L/\tilde{\xi}}) \pmod{1} ,
\end{align}
we can extract the difference in topological spin $\Delta h = h_{\Omega^2} - h_{\Omega^1}$.
As shown in Fig.~\ref{deltas}, we get $\Delta h \approx -0.21$ for a number of points in phase space.
This is consistent with the identification $\Omega^1 = \psi$, $\Omega^2 = \spinon$, as $h_\spinon - h_\psi = \frac{5}{16} - \frac{1}{2}  = -0.1875$ in the iPf phase. We attribute the difference between the observed and expected values to finite-size and finite-$\chi$ systematic errors.
(Note that $h_{\mathds{1}} = 0$, and thus we can conclude $\Omega^1 \neq \mathds{1}$.)

Further support for our identification of ground states can be found in the entanglement spectrum.
We first give the theoretical orbital entanglement spectra for the ground states of the iPf phase, which depends on both the ground state $\ket{a}$ and the charge across the entanglement cut.
(Note that so far in this work, we have given the entanglement spectra for only one value of electric charge crossing the entanglement cut, we chose the value of charge which has the lowest lying entanglement states.  Henceforth we will be explicit about the charges.)
For any of the nine MES and fixed charge $(Q,S^z)$ across the entanglement cut, the entanglement spectra counting follows one of three possible sequences.
\begin{align}\begin{split}
	s_1: 1, 2, 6, 13, \dots,
\\	s_\sigma: 1, 3, 8, 19, \dots,
\\	s_\chi: 1, 3, 8, 18, \dots.
\end{split}\end{align}
For state $\ket{\mathds{1}}$, the entanglement spectrum follows the $s_1$ sequence for even $Q$, and $s_\chi$ sequence for odd $Q$.
For state $\ket{\psi}$, the spectrum follows $s_\chi$ and $s_1$ for even and odd $Q$ respectively.
For state $\ket{\spinon}$, the entanglement spectrum always follows $s_\sigma$.


\begin{figure}[tb]
	\includegraphics[width=\linewidth]{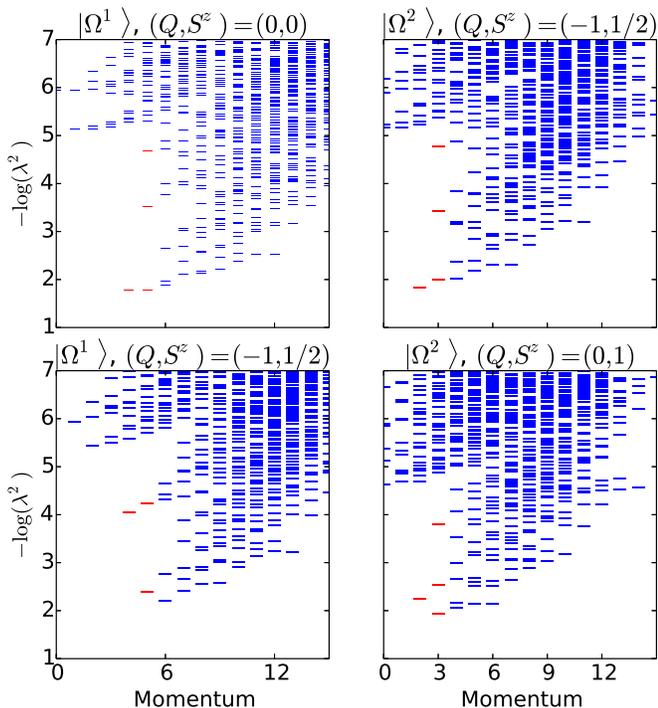}
	\caption{(Color online)
		Entanglement spectra for the putative iPf state.
		The left two panels show the entanglement spectra for the $\ket{\Omega^1}$ state, for the charge sectors with the lowest-lying and second lowest-lying entanglement states, the counting for these states is $1,3,...$ and $1,2,...$, as expected if $\ket{\Omega^1}=\ket{\psi}$.
		The right panel shows spectra for the $\ket{\Omega^2}$ state.
		There are two degenerate charge sectors with lowest lying states.
		Here we show one example from each of the two sectors with the lowest-lying entanglement states, and we find counting of $1,3,\dots$ in both, as expected if $\ket{\Omega^2}=\ket{\spinon}$.
	}
	\label{pfaffspectra}
\end{figure}

We attempt to match up the low-lying states of the ground states $\ket{\Omega^1}$ and $\ket{\Omega^2}$ to those expected for the iPf phase, shown in Fig.~\ref{pfaffspectra}.
Typically one defines the ``low-lying" entanglement states as those below the ``entanglement gap", which is a window devoid of states as the circumference is increased.
In practice, at finite system size we observe multiple regions without states which could be called the entanglement gap, which makes it difficult to specify which levels should be counted.
We have highlighted in the figures to indicate the presumed counting of iPf, but in full honesty other assignments are possible.
The left panels shows the entanglement spectra for $\ket{\Omega^1}$, with $(Q,S^z) = (0,0)$ (top) and $(Q,S^z) = (-1,\frac12)$ (bottom).
Assuming states with $-\log(\lambda^2) < 5$ are the low-lying states, we observe the counting $1,3,\dots$ and $1,2,\dots$ for the two charge sectors, respectively.
This suggests that $\ket{\Omega^1} = \ket{\psi}$, consistent with the momentum polarization data above.
On the right panels, we showed the entanglement spectra for $\ket{\Omega^2}$ (for the same charges), which seem to indicate the counting $1,3,\dots$ regardless of charge, also consistent with the identification $\ket{\Omega^2} = \ket{\spinon}$.


In summary, the well-established spin-charge separation shown in Fig.~\ref{charge} rules out all currently proposed wavefunctions besides the intralayer-Pfaffian and iPf phase.
The entanglement properties and the overlaps, though not conclusive, are incompatible with the intralayer-Pfaffian state, but appear consistent with the iPf state.

\section{Conclusion}\label{sec:conclusions}

In this work, we use iDMRG and exact diagonalization techniques to study a bilayer quantum Hall system with filling $1/3$ in each layer. We find a phase diagram in terms of the experimentally accessible parameters: layer separation, interlayer tunnelling, and layer width. We find three different phases: a phase with decoupled layers, a bilayer-spin singlet phase, and a bilayer-symmetric phase. We confirm the nature of these phases and study the transitions between them.

We also explore the phase diagram for Coulomb interaction with modified short-range components ($V_0$ and $V_1$). We find a non-Abelian phase over a wide region of parameter space.
This phase has anyons which carry spin $1/2$ and no charge.
This observation, coupled with a study of additional entanglement properties and wavefunction overlaps, leads us to conclude that the non-Abelian phase is interlayer-Pfaffian.
Our data for the non-Abelian phase is inconsistent with all other known non-Abelian candidates.
However, it is possible that a novel non-Abelian state could be constructed that can reproduce our results.

Although it is experimentally not feasible to directly modify a given pseudopotential, there are many realistic ways to change the Coulomb interaction in a quantum Hall system, such as varying the chemical potential to place the system in a higher Landau level, introducing Landau level mixing, tilting the magnetic field, or screening the Coulomb potential. 
Since all of these perturbations can ultimately be expanded in terms of $V_\alpha$'s, the phase diagram we find here may be helpful in guiding such studies towards a realization of the non-Abelian phase. 

\acknowledgments
We acknowledge helpful conversations with Eddy Ardonne, Maissam Barkeshli, Parsa Bonderson, and Xiaoliang Qi.
We are particular grateful to Jim Eisenstein, without whom this work would not have been started.
S.G. is supported by National Science Engineering Research Council (NSERC) of Canada.
M.Z. is indebted to the David \& Lucile Packard Foundation.
Z.P. acknowledges support by DOE grant DE-SC0002140.
Research at Perimeter Institute is supported by the Government of Canada through Industry Canada and by the Province of Ontario through the Ministry of Economic Development \& Innovation. 
R.M. acknowledges the Sherman Fairchild Foundation for support.


\bibliography{bib23}

\end{document}